\documentclass[conference]{IEEEtran}
\IEEEoverridecommandlockouts

\usepackage{cite}
\usepackage{amsmath}
\usepackage{amssymb}
\usepackage{amsfonts}
\usepackage{algorithmic}
\usepackage{graphicx}
\usepackage{textcomp}
\usepackage{soul}
\usepackage{xcolor}
\usepackage{bm}
\usepackage[T1]{fontenc}

\def\BibTeX{{\rm B\kern-.05em{\sc i\kern-.025em b}\kern-.08em
    T\kern-.1667em\lower.7ex\hbox{E}\kern-.125emX}}

\begin{document}

\title{On Noise Resiliency of Neuromorphic Inferential Communication in Microgrids \\
}

\author{
\IEEEauthorblockN{Yubo Song}
\IEEEauthorblockA{
\textit{Department of Energy} \\
\textit{Aalborg University} \\
Aalborg, Denmark \\
yuboso@energy.aau.dk}
\and
\IEEEauthorblockN{Subham Sahoo}
\IEEEauthorblockA{
\textit{Department of Energy} \\
\textit{Aalborg University} \\
Aalborg, Denmark \\
sssa@energy.aau.dk}
\and
\IEEEauthorblockN{Xiaoguang Diao}
\IEEEauthorblockA{
\textit{Hubei University of} \\
\textit{Technology} \\
Wuhan, China \\
dxg202020@163.com}
}

\maketitle

\begin{abstract}
Neuromorphic computing leveraging spiking neural network has emerged as a promising solution to tackle the security and reliability challenges with the conventional cyber-physical infrastructure of microgrids. Its event-driven paradigm facilitates promising prospect in resilient and energy-efficient coordination among power electronic converters. However, different from biological neurons that are focused in the literature, microgrids exhibit distinct architectures and features, implying potentially diverse adaptability in its capabilities to dismiss information transfer, which remains largely unrevealed. One of the biggest drawbacks in the information transfer theory is the impact of noise in the signaling accuracy. Hence, this article hereby explores the noise resiliency of neuromorphic inferential communication in microgrids through case studies and underlines potential challenges and solutions as extensions beyond the results, thus offering insights for its implementation in real-world scenarios.
\end{abstract}


\section{Introduction}

Microgrids are serving as critical interfaces for emerging distributed energy resources (DERs), whereas challenges have been encountered in effectively coordinating the DERs to optimize energy efficiency as well as costs. To this end, hierarchical control and communication techniques are introduced as prospect solutions. Nevertheless, latency \cite{yan2023latency}, communication link failures \cite{wu2022line}, and cybersecurity issues \cite{zhiyi2017cybersec} arising out of the highly unreliable communication layer in conventional cyber-physical infrastructure induces stability and blackout threats to the entire system \cite{dong2019stability}.

The concept of \textit{Talkative Power Communication} (TPC) emerges as a practical solution, where the power transmission lines are leveraged as physical information channels and information is encoded through digital modulation techniques \cite{ang2016powertalk, he2020nature, marco2023tpc}. By co-transferring power and information simultaneously, TPC has shown advantages in reducing the cost of additional physical information channels and improving the resiliency of microgrid systems. To this end, the scalability of TPC is still an aspect that needs further investigation.

The development of next-generation artificial intelligence (AI) techniques further address this issue by implementing the operation and coordination without any communication channels. The information sampled remotely, which is physically constrained by the system architecture, can be estimated by detecting the dynamics observed locally. Besides, as compared to the traditional second-generation neural networks, spiking neural network (SNN) stands out as a pioneering event-based data-driven learning technique inspired by the intricate biological neural-synaptic framework \cite{roy2019nature}. As neuromorphic computing leveraging SNN has emerged for sensing and communication networks \cite{chen2022neuromorphic}, it has been first deployed in the distributive control of microgrids just using power flows as a means of coordination \cite{sahoo2024nsc}, heralding neuromorphic communication that can infer remote information locally in microgrids notably enhances reliable system performance.

Meanwhile, neuromorphic computation is largely challenged by its accuracy and versatility, which often plays a big impact due to the noise in the measurement signals. In the literature, there has been modeling and discussion on the noise resiliency of biological and artificial neuron models, whereas the neuromorphic communication for microgrids, which we formalized in our recent efforts \cite{sahoo2024nsc} and \cite{spiket}, has not primarily considered those aspects. Considering the scaling difference between a neuromorphic circuit and a microgrid in terms of voltage, power level and electrical applications, the inferential capabilities of communication in microgrids remains ambiguous, thus becoming a prime motivation behind the study in this paper.

Hence, we scrutinize the noise resiliency in each converter's sensors that ultimately affect the neuromorphic inferential process and its communication with other converters. The rest of the article is organized as following: Section II introduces how the neuromorphic inferential infrastructure is employed in microgrids. Section III provides a comprehensive validation behind noise resiliency under simulation and experimental environment. Going beyond the results, Section IV provides further insights in the role of noise in neuromorphic communication. Section V concludes the entire article and summarizes future perspective of extensions on this topic.


\section{Neuromorphic Inferential Communication in Microgrids}

\subsection{Neuromorphic Algorithms and Spiking Neural Network}

\begin{figure}[t]
    \centering
    \includegraphics[scale=1.0]{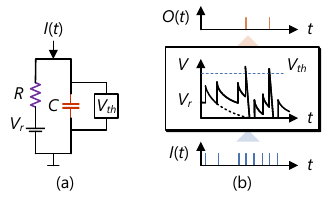}
    \caption{Concept of the the Leaky Integrate-and-Fire (LIF) neuron model: (a) equivalent \textit{RC} circuit, and (b) input and output spikes driven by the model, where the $V_\mathrm{th}$ based criteria is employed.}
    \label{fig_RC}
\end{figure}

The concept of neuromorphic computing is inspired by the physics of information transmission of neurons in our brains \cite{yu2017neuromorphic}. In nature, the neurons are interconnected via synapses, which are called pre- and post-synaptic neurons. The neurons fire only when there is a stimuli or \textit{event}, namely the \textit{event-driven} paradigm. As per computational neuroscience, the electrical behavior of membrane separating two neurons and the ion passages for information exchange functions like capacitor $C$ and resistor $R$, respectively, which is eventually modeled into an \textit{RC} circuit, as shown in \ref{fig_RC}(a).

In this \textit{RC} circuit, the membrane potential $V_\mathrm{mem}$, which corresponds to the capacitor voltage, will be changing over time following the input current $I(t)$ injected to the neuron. As a result, the \textit{RC} circuit will be charged with every instance of a spike, otherwise discharges to a decaying potential. Due to this property, this model is consequently named as the \textit{Leaky Integrate-and-Fire} (LIF) model \cite{yu2017neuromorphic}. In LIF model, the membrane potential follows the \textit{RC} dynamics in time domain:
\begin{equation}
    I(t) = \frac{V_\mathrm{mem}(t) - V_r}{R} + C\frac{\mathrm{d}V_\mathrm{mem}}{\mathrm{d}t}
\end{equation}

By defining the decay time constant $\tau_m$ = $RC$ as the \textit{leaky integrator}, the following holds:
\begin{equation}
\label{eq_mem}
    \tau_m \frac{\mathrm{d}V_\mathrm{mem}}{\mathrm{d}t} = -[V_\mathrm{mem}(t) - V_\mathrm{r}] + \cfrac{I(t)}{g}
\end{equation}
where, $g$ is the leaky conductance.

Subsequently, the event is passed through the synapse when the membrane potential $V_\mathrm{mem}$ surpasses a threshold $V_\mathrm{th}$ of the post-synaptic neuron, and the output spikes are generated as the triggered output events, as shown in Fig. 1(b). While the charging and discharging processes are repeated over time, the neuronal dynamics turn out to be an integration process. By extending this architecture to a neural network, the event-driven \textit{Spiking Neural Network} (SNN) is formulated, establishing the basis for neuromorphic computation and communication.

\begin{figure}[t]
    \centering
    \includegraphics[width=0.9\linewidth]{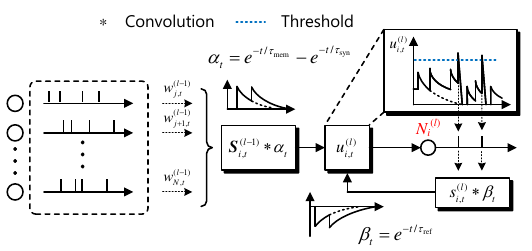}
    \caption{Spike response model (SRM) for simulation of the leaky and integrate fire (LIF) neuron.}
    \label{fig_snn}
\end{figure}

The spike response model (SRM) in SNN is depicted in Fig. \ref{fig_snn} \cite{sahoo2024nsc}, which is widely used to represent the behaviors of biological neurons. The membrane potential $u^{(l)}_{i,\:t}$ of the neuron $N^{(l)}_{i}$ increases when there is a spike, and decays in an exponential trend, which is modeled as a synaptic filter:
\begin{equation}
    \alpha_t = e^{-t/\tau_\mathrm{m}} - e^{-t/\tau_\mathrm{syn}}
\end{equation}
with the decay process modeled as a feedback filter:
\begin{equation}
    \beta_t = - e^{-t/\tau_\mathrm{ref}}
\end{equation}
where, $\tau_\mathrm{m}$, $\tau_\mathrm{syn}$ and $\tau_\mathrm{ref}$ are finite positive constants.

The membrane potential $u_{i,\:l}$ are thereby given as:
\begin{equation}
    u^{(l)}_{i,\:t} = \sum_{j=1}^{N}w^{(l-1)}_{j}\cdot(\alpha_t*s^{(l-1)}_{j,\:t}) + \beta_t*s^{(l)}_{i,\:t}
\end{equation}
where, $*$ is the convolution operator.

The spikes $s^{(l)}_{i,\:t}$ is generated by the neuron $N^{(l)}_{i}$ (from Layer $i$) when the threshold $U_\mathrm{thr}$ is hit:
\begin{equation}
    s^{(l)}_{i,\:t} = H(u^{(l)}_{i,\:t}-U_\mathrm{thr})
\end{equation}
where, $H(\cdot)$ is the Heaviside step function:
\begin{equation}
    H(x)=\left\{
    \begin{array}{rcl}
    1 & & {x > 0}\\
    0 & & {x \leq 0}
    \end{array} \right.
\end{equation}

\subsection{Neuromorphic Inferential Communication in Microgrids}

\begin{figure}[t]
    \centering
    \includegraphics[width=0.9\linewidth]{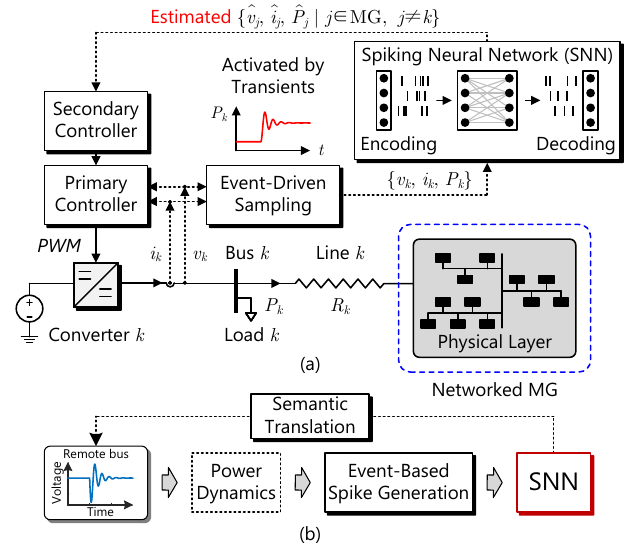}
    \caption{Architecture of a neuromorphic inferential communication in DC microgrid: (a) infrastructure of SNN based control framework for converter $k$, and (b) flowchart of neuromorphic inferential state estimation.}
    \label{fig_system}
\end{figure}

An integrated neuromorphic infrastructure based control and coordination philosophy is illustrated in Fig. \ref{fig_system}, where a DC microgrid is taken as an example \cite{sahoo2024nsc}. Different from the conventional cyber-physical infrastructure, it transcends the reliance on physical tie-line power flows for the SNNs at each node/edge to infer remote information, which is named as neuromorphic communication. In Fig. \ref{fig_system}, the locally sampled voltage and current are harnessed for training the SNN based on a sparse event-driven fashion. This means spikes based data is collected only during dynamic conditions. With this criteria, remote information can be inferred from the spatio-temporal patterns in spikes, which is aligned with the \textit{publish-subscribe} information theoretic protocol \cite{sahoo2024nsc} and facilitating optimal regulation of the power flow based on the secondary control objectives.

Considering the DC microgrid case where distributed energy resources (DERs) share the load current equally. Furthermore, the local voltage commands for DER $k$ are generated by the following equation \cite{rath2023blockchain}:
\begin{equation}
    v^*_k(t) = v_{ref,\:k}(t) + \delta v^{\mathrm{I}}_k(t) + \delta v^{\mathrm{II}}_k(t)
\end{equation}
where, the regulation terms $\delta v^{\mathrm{I}}_k$ and $\delta v^{\mathrm{II}}_k$ are derived from primary voltage observer and secondary power sharing objective, respectively. As both regulators are implemented by PI controller, the error terms are defined as:
\begin{subequations}
\begin{align}
    e_{v_k}^\mathrm{I} &= V_\mathrm{ref} - \bar{v}_k(t) \\
    e_{v_k}^\mathrm{II} &= \sum_{j\in N_k}a_{kj}(P_j(t)-P_k(t))
\end{align}
\end{subequations}
where, $V_\mathrm{ref}$ is the reference of average voltage in the system, $N_k$ represents the set of nodes that are adjacent to DER $k$, and $\bar{v}_k(t)$ is the average voltage observed from bus $k$:
\begin{equation}
    \bar{v}_k(t) = v_k(t)+\int_{0}^{t}\sum_{j\in N_k}a_{kj}(\bar{v}_j(\tau)-\bar{v}_k(\tau))\mathrm{d}\tau
\end{equation}

In this case, the SNN-dominated infrastructure in Fig. \ref{fig_system}(a) is subsequently used to estimate the remote data of converters \cite{sahoo2024nsc}. The events to be captured are essentially the power dynamics, which are determined by the physical characteristics of the filters. Considering the capacitor current $i^C_k$, the inductor voltage $v^L_k$, and corresponding equivalent voltage gain of each DC-DC converter $d_k$:
\begin{subequations}
\begin{align}
    i^\mathrm{C}_k (t) &= C_k\cfrac{\mathrm{d}v_k}{\mathrm{d}t} = i_k (t) - d_k i^\mathrm{in}_k (t) \\
    v^\mathrm{L}_k (t) &= L_k\cfrac{\mathrm{d}i_k}{\mathrm{d}t} = v_k (t) - d_k v^\mathrm{in}_k (t)
\end{align}
\end{subequations}
The input events of the SNN are triggered by the local error signals as:
\begin{subequations}
\begin{align}
\label{eq_evt_vi}
    \Omega_\mathrm{i} (t) &= v^\mathrm{L}_k (t) - e^i_k (t) \\
    \Omega_\mathrm{v} (t) &= i^\mathrm{C}_k (t) - e^v_k (t)
\end{align}
\end{subequations}

The remote state estimation in the system can be similarly formulated, nevertheless in the vector form considering all adjacent agents:
\begin{equation}
    \mathbf{C}\mathbf{\dot{V}}(t) = \mathbf{JI_{flow}}(t) - \mathbf{d i^{in}}(t)
\end{equation}
where, $\mathbf{J}$ is a matrix with binary values indicating the physical connection of two nodes. $\mathbf{C}, \mathbf{V}, \mathbf{d}$ and $\mathbf{i^{in}}$ denote the capacitance, voltage, voltage gain and the input current of the converters, respectively. $\mathbf{I_{flow}}$ depicts the tie-line flow currents into the connected lines from the converters.

The output events are generated to reflect the global dynamics of the entire system:
\begin{equation}
\label{eq_evt_o}
    \Omega_\mathrm{o}(t) = C_k \dot{v}_k(t) - \dot{I}^k_\mathrm{flow}(t)
\end{equation}

An event is said to be triggered, when any one of the three indicators exceeds a given threshold:
\begin{equation}
    || \Omega_\mathrm{v}(t) || > \sigma^\mathrm{V}_\mathrm{th}, \quad ||\Omega_\mathrm{i}(t)|| > \sigma^\mathrm{I}_\mathrm{th}, \quad ||\Omega_\mathrm{o}(t)|| > \sigma^\mathrm{o}_\mathrm{th}
\end{equation}
where, $\sigma^\mathrm{V}_\mathrm{th}$, $\sigma^\mathrm{I}_\mathrm{th}$ and $\sigma^\mathrm{o}_\mathrm{th}$ are the corresponding thresholds for (12a), (12b) and (14), respectively. The triggered events are then encoded and translated into input spikes for SNN to estimate the remote measurements, and decoded for subsequent controllers. With this event-driven paradigm, SNN shows great advantage in reducing the energy consumption and thus has promising prospect of energy-efficient operation of microgrids. 

\section{Performance Validation}

As neuromorphic inferential communication is an event-driven paradigm, its performance is closely dependent on event capturing, which is subject to sampling noise in practice. To this end, this section aims to inspect this aspect in terms of microgrid applications, and to draw preliminary conclusions on the solutions to tackle this issue.

\begin{figure}[t]
    \centering
    \includegraphics[width=0.9\linewidth]{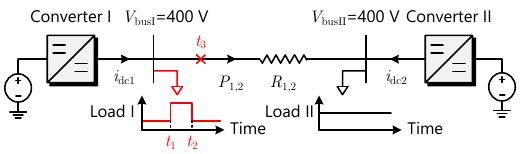}
    \caption{A two-bus DC microgrid as the test case.}
    \label{fig_testcase}
\end{figure}

A two-bus DC microgrid shown in Fig. \ref{fig_testcase} is first selected as a simple test case, where the aforementioned control objectives are accommodated. The neuromorphic infrastructure are implemented via Python-Simulink co-simulation, and several transient scenarios are tested, as marked in Fig. \ref{fig_testcase}. The system and control parameters can be found in Appendix.

\begin{figure}[t]
    \centering
    \includegraphics[width=0.8\linewidth]{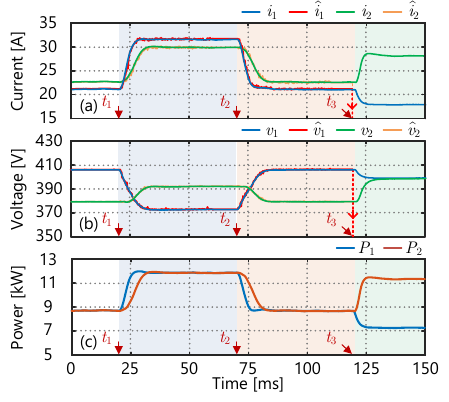}
    \caption{Performance of microgrid governed by neuromorphic controllers: comparison of (a) estimated and measured currents, (b) estimated and measured voltages, and (c) estimated and measured powers.}
    \label{fig_snnvi}
\end{figure}

The performances without considering sampling noise are demonstrated in Fig. \ref{fig_snnvi}. The estimated values follow the sampled values in satisfactory accuracy with the fundamental control objectives fulfilled, and they keep in accordance during the transients. Based on the results, the rationality of the study case can be confirmed, as well as the feasibility of neuromorphic communication in ideal cases.

\subsection{Validation of System Performance by Simulation}

\begin{figure}[t]
    \centering
    \includegraphics[width=1.0\linewidth]{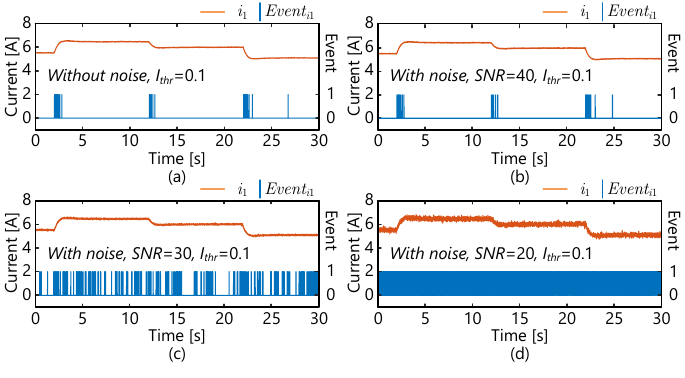}
    \caption{Event capturing with different noise added to the sampled signal: (a) without noise, $I_\mathrm{thr}$ = 0.1, (b) with AWGN, SNR = 40, $I_\mathrm{thr}$ = 0.1, (c) with AWGN, SNR = 30, $I_\mathrm{thr}$ = 0.1, (d) with AWGN, SNR = 20, $I_\mathrm{thr}$ = 0.1.}
    \label{fig_eventcap_snr}
\end{figure}
\begin{figure}[t]
    \centering
    \includegraphics[width=1.0\linewidth]{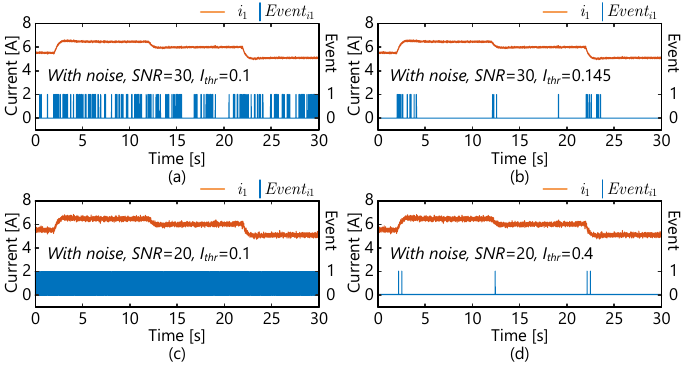}
    \caption{Event capturing with different threshold applied: (a) with AWGN, SNR = 30, $I_\mathrm{thr}$ = 0.1, (b) with AWGN, SNR = 30, $I_\mathrm{thr}$ = 0.145, (c) with AWGN, SNR = 20, $I_\mathrm{thr}$ = 0.1, (d) with AWGN, SNR = 20, $I_\mathrm{thr}$ = 0.4.}
    \label{fig_eventcap_thr}
\end{figure}

Noise exerts its influences on the sampling as well as information theoretic learning with neuromorphic computation. Hence, we further extend this consideration by testing under the influences of the basic additive white Gaussian noise (AWGN). An initial assessment on the event synthesis using (\ref{eq_evt_vi}) and (\ref{eq_evt_o}) is conducted. The results are illustrated in Fig. \ref{fig_eventcap_snr}. In an ideal case without noise in Fig.\ref{fig_eventcap_snr}(a), the events precisely align with the transients, whereas noise may cause the events to be over-captured, as evident in Fig.\ref{fig_eventcap_snr}(b)-(d). In this case, when the signal-to-noise ratio SNR is set to $\leq$ 30, the transients cannot be correctly detected as events with the current threshold, leading to an over-execution of SNN.

Apart from additional noise filters, a practical solution could involve the modification of the threshold as demonstrated in Fig. \ref{fig_eventcap_thr}. The accuracy of event capturing is restored by lifting the threshold of event capturing to counteract the decrease of SNR, and lower SNR corresponds to more increase of threshold. In Fig. \ref{fig_eventcap_thr}, the events can be captured when increasing the threshold $I_\mathrm{thr}$ to 0.4, but it should be also noted that this method is essentially be constrained by the ability to distinguish the events from the disturbances of noise.

\subsection{Validation of Noise Resiliency in Training and Execution Process with Simulation Data}

\begin{figure}[t]
    \centering
    \includegraphics[width=1.0\linewidth]{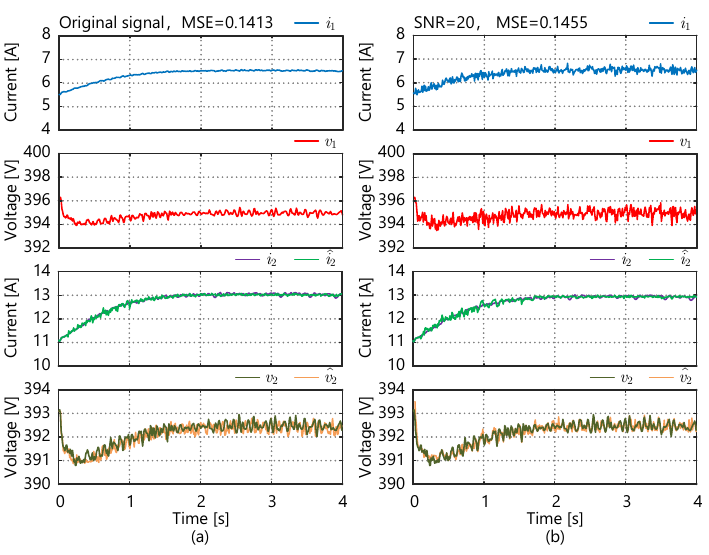}
    \caption{Results regarding noise in the training process, where the currents of Converter I and II are input and output data, respectively: (a) input without noise, (b) input with an AWGN of SNR=20.}
    \label{fig_noise1_0}
\end{figure}
\begin{figure}[t]
    \centering
    \includegraphics[width=1.0\linewidth]{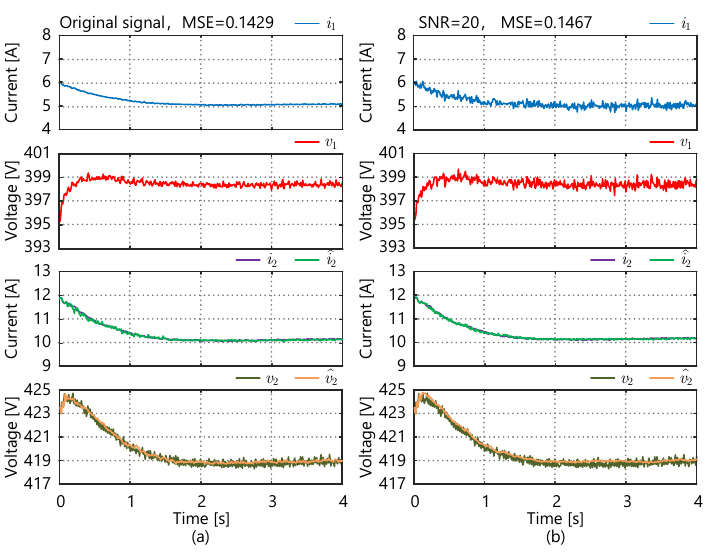}
    \caption{Results regarding noise in the execution process, where the currents of Converter I and II are input and output data, respectively: (a) input without noise, (b) input with an AWGN of SNR=20.}
    \label{fig_noise2_0}
\end{figure}

Furthermore, we validate the performances under the influence of noise during online training and execution processes, with results presented in Fig. \ref{fig_noise1_0} and Fig. \ref{fig_noise2_0}, respectively. The sampled signals exhibit differences in the ripples, while there is less influence on the estimated signals, due to the relatively low bandwidth of the control loops compared to the noise. The mean squared errors (MSE) of the signals are also measured, implying the that the feasible noise tolerance of neuromorphic communication for microgrids using SNN. 

\subsection{Validation of Spike Synthesis with Experimental Data}

\begin{figure}[t]
    \centering
    \includegraphics[width=0.9\linewidth]{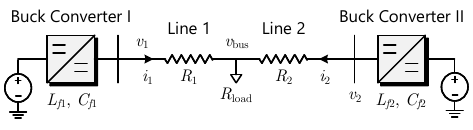}
    \caption{A two-bus DC microgrid as the test case.}
    \label{fig_expsys}
\end{figure}
\begin{figure}[t]
    \centering
    \includegraphics[width=0.75\linewidth]{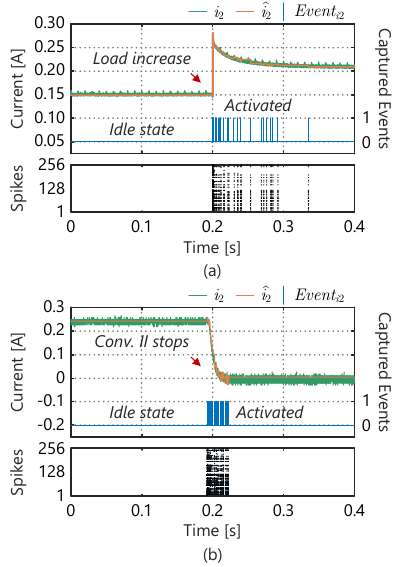}
    \caption{Validation of the noise resiliency by using the data obtained from a down-scaled experimental system.}
    \label{fig_exp}
\end{figure}

The noise resiliency of neuromorphic inferential communication is also validated with the data obtained from a down-scaled experimental setup. The system under test is configured as Fig. \ref{fig_expsys}, which is also a DC microgrid consisting of two DC-DC buck converters, tied to a resistive load. The current from Converter 2 is sampled, and two scenarios are validated, including a load increase and an outage of the Converter 2.

The test results are presented in Fig. \ref{fig_exp}. Though noise naturally exists in the measurements across the experimental platform, spikes are synthesized right after the transient. Further, the estimated current $\hat{i}_2$ seen from Converter 1 can follow the sampled value to an acceptable extent. This can significantly justify that neuromorphic communication has an acceptable level of noise resiliency in microgrid applications and its feasibility in practice.

\section{Further Perspectives on the Noise Resiliency}

Going beyond the preliminary results, we elucidate in this section further perspectives on the noise resiliency in terms of the physics behind neuromorphic computation. As the infrastructure is governed by the LIF neuron model, some of its characteristics functions distinctly as compared with conventional neurons based on simple summation functions. To this end, we tend to extend the discussions to the following aspects:

\subsubsection{Low Pass Filter in the LIF neuron model}
From (\ref{eq_mem}), there is a low-pass filter (LPF) naturally inside the LIF neuron model from the \textit{RC} equivalent circuit, which provides frequency-domain attenuation for the input to some extent. In frequency domain, the LPF can be modeled as:
\begin{equation}
\label{eq_mem_inv}
    \frac{V_\mathrm{mem}(s)}{I(s)} = \cfrac{1}{g \cdot (1+\tau_m s)}
\end{equation}
The natural cut-off frequency of this LPF is determined by the \textit{RC} time constant $\tau_m$:
\begin{equation}
    \omega_c = \cfrac{1}{\tau_m}
\end{equation}
of which, according to \cite{dayan2005neuroscience}, an exemplary value is around 10-100 ms in real neurons, equivalent to a cut-off frequency of 1.6-16 Hz, while in SNN, it is more flexible.

This inherent frequency depicts how fast the neuron restores back to steady-state after encountering a spike. Hence, smaller $\tau_m$ may lead to larger number of output spikes, and higher achievable temporal resolution for congested events.

\subsubsection{Noise modeling for LIF neurons} 
The noise modeling for LIF neurons has been preliminarily discussed in \cite{naud2014neurodyn}, where the real neurons are focused. Additive noise is considered in the LIF model (\ref{eq_mem}), namely:
\begin{equation}
\label{eq_mem_n}
    \tau_m \frac{\mathrm{d}V_\mathrm{mem}}{\mathrm{d}t} = -[V_\mathrm{mem}(t) - V_\mathrm{r}] + \cfrac{I(t)}{g} + \xi(t)
\end{equation}
where, $\xi(t)$ represent the noise, and for Gaussian noise $\langle \xi(t) \rangle = 0$. By taking the integration of (\ref{eq_mem_n}) over time, the noise will drift the membrane potential away from the reference trajectory, thus leading to mis-triggered events.

Meanwhile, it has also been pointed out in \cite{naud} that an average of the interspike interval $\langle \mathrm{ISI} \rangle$ need to be taken when accounting for the noise. In this way, low synaptic current contribution is enhanced for a linear profile in Siegert equation \cite{naud}, since the additive noise in return contributes to linearization of the input-output function.

However, it should also be noted that the above theories are established based on the neurons in our brains, of which the voltage is within several mV. In microgrid applications, the voltage ranges from several to hundreds of volts, and it will be one of our future focuses to extend on this point and accommodate the theory to practice.

\subsubsection{Bandwidth of controllers}
Different from the biological neurons, there are additional control loops in microgrid applications, like the primary and secondary observers in the study cases of this article. The bandwidth of the controllers, which locates around kHz in most cases, will provide additional resiliency against noise. In the future, we tend to perform frequency-domain analysis and inspect the power spectrum, so as to develop the optimized design of controllers from this perspective.

\subsubsection{Impacts on Hebbian learning}
The concept of Hebbian learning has been emphasized in \cite{spiket}, which is a distinct feature of SNN, allowing online learning via varied threshold $V_\mathrm{th}$. Nevertheless, as noise may cause the membrane potential $V_\mathrm{mem}$ to mis-hit the threshold, the convergence of Hebbian learning may be affected, which should also be a future perspective that can benefit the energy efficiency of neuromorphic applications in microgrids.

\section{Conclusions and Future Scope of Work}

This article has delved into noise resiliency of neuromorphic inferential communication in microgrids, shedding light on the potential challenges and solutions posed by noise through test cases. We have showcased the noise resiliency of SNN, implying promising applications of energy-efficient edge computing in practice.

Besides, a few perspectives are provided in relation to both the physics behind neuromorphic computing and its corresponding modeling and deployment challenges in microgrid applications. As a future scope of work, we aim to conduct more analysis in the frequency domain, to characterize the power spectrum of the noise and formalize a more noise-resilient and energy-efficient framework for brain-inspired coordination in microgrids.

\appendix
\subsection{SNN Parameters -- Simulation Studies}
Number of hidden layers = 2, Number of neurons in encoding and hidden layer = 256, Number of neurons in decoding layer = 4, $\sigma^{V}_{th}$ = 0.01, $\sigma^{I}_{th}$ = 0.002, $\sigma^{I}_{th}$ = 0.0039.\par
Dataset dimensions for the inputs and outputs in Case I and III : $\mathbb{D}_{in,800\times 4}\:=\:\{\bm{v}_{i,800\times 1},\:\bm{i}_{i,800\times 1},\:\bm{\dot{v}}_{i,800\times 1},\:\bm{\dot{i}}_{i,800\times 1}\}$, {\mbox{$\mathbb{D}_{out,800\times 2}\:=\:\{\bm{\hat{v}}^{(i)}_{j,800\times 1},\:\bm{\hat{i}}^{(i)}_{j,800\times 1},\}$}}, 
$\forall \ i,j \in \{1,2\}, i\neq j$.

Dataset dimensions for the inputs and outputs in Case II and IV : $\mathbb{D}_{in,800\times 4}\:=\:\{\bm{v}_{i,800\times 1},\:\bm{i}_{i,800\times 1},\:\bm{\dot{v}}_{i,800\times 1},\:\bm{\dot{i}}_{i,800\times 1}\}$, {\mbox{$\mathbb{D}_{out,800\times 4}\:=\:\{\bm{\hat{v}}^{(i)}_{j,800\times 1},\:\bm{\hat{i}}^{(i)}_{j,800\times 1},\:\bm{\hat{v}}^{(i)}_{k,800\times 1},\:\bm{\hat{i}}^{(i)}_{k,800\times 1}\}$}}, 
$\forall \ i, j, k \in \{1, 2, 3\}, i\neq j\neq k$.
\subsection{SNN Parameters -- Experimental Studies}
Number of hidden layers = 2, Number of neurons in encoding and hidden layer = 64, Number of neurons in decoding layer = 4, $\sigma^{V}_{th}$ = 0.41, $\sigma^{I}_{th}$ = 0.0063, $\sigma^{I}_{th}$ = 0.024.\par
Dataset dimensions for the inputs and outputs: $\mathbb{D}_{in,4000\times 4}\:=\:\{\bm{v}_{i,4000\times 1},\:\bm{i}_{i,4000\times 1},\:\bm{\dot{v}}_{i,4000\times 1},\:\bm{\dot{i}}_{i,4000\times 1}\}$, {\mbox{$\mathbb{D}_{out,4000\times 2}\:=\:\{\bm{\hat{v}}^{(i)}_{j,4000\times 1},\:\bm{\hat{i}}^{(i)}_{j,800\times 1},\}$}}, 
$\forall \ i,j \in \{1,2\}, i\neq j$.
\bibliographystyle{IEEEtran}
\bibliography{bibligraphy}

\end{document}